\documentclass[aps,prc,twocolumn,floatfix,showpacs,a4paper,nofootinbib]{revtex4-1}
\usepackage{graphicx}
\usepackage{epsfig}
\usepackage{amssymb,latexsym,amsmath}
\usepackage{soul}
\usepackage{color}

\begin{document}

\title{Longitudinal fluctuations of the center of mass of the participants in heavy-ion collisions}
\author{V. Vovchenko$^1$, D. Anchishkin$^2$, and L.P. Csernai$^3$}
\affiliation{
$^1$Taras Shevchenko Kiev National University, Kiev 03022, Ukraine}
\affiliation{
$^2$Bogolyubov Institute for Theoretical Physics, Kiev 03680, Ukraine}
\affiliation{
$^3$Institute for Physics and Technology, University of Bergen,
5007 Bergen, Norway}
\date{\today}

\begin{abstract}
A model for computing
the probability density of event-by-event participant
center-of-mass rapidity $y^{c.m.}$
is presented.
The evaluations of the $y^{c.m.}$ distribution are
performed for different collision energies and different
centralities.
We show that for certain conditions the
rapidity distribution is described by a Gaussian
with a variance determined mostly by the collision
centrality.
It is found that the
width of the $y^{c.m.}$ distribution
increases strongly for more peripheral
collisions, while it depends weakly on the collision energy.
Other theoretical estimates of rapidity distribution
are presented and questions of interaction and separation
between spectators and participants
are discussed.
\end{abstract}


\maketitle

\section{Introduction}

To describe the many-particle interacting system created in heavy-ion collisions
(participant system) different
models such as hydrodynamics or kinetic transport models are used. Along
with the participants there are also spectators,
which are nucleons emerged from the colliding nuclei that do not take
part in any
reaction with other nucleons
during the collision process and move with their initial
momenta. The number of spectators from each of the nuclei changes
event-by-event (EbE)
and, due to this
fluctuation,
the center-of-mass of participant system does not
coincide with the collider center-of-mass system (c.m.s), i.e. the participant c.m.
rapidity, $y^{c.m.}$, may be non-zero in a particular event.
The EbE fluctuations
of $y^{c.m.}$ can be especially significant in peripheral collisions
when impact parameter of colliding nuclei is large and the mass of the
spectators is essential.

When comparing different observables which depend on rapidity,
for instance collective flow
(calculated e.g. in a hydrodynamical model) with experimental measurements,
it could be important to account for participant c.m. fluctuations,
which may influence the results~\cite{Csernai1,Csernai2,Csernai3}.
Possible influence of EbE longitudinal
fireball density fluctuations on measurable two-particle rapidity correlation
function was recently studied in \cite{BT13}.

In the present work a simplified model for the calculation of
the EbE $y^{c.m.}$ distribution is presented, and center-of-mass
rapidity fluctuations are discussed.

\section{The Model}
\label{sec:model}

\subsection{Participant rapidity from spectators}

\begin{figure}
\begin{center}
\includegraphics[width=0.3\textwidth]{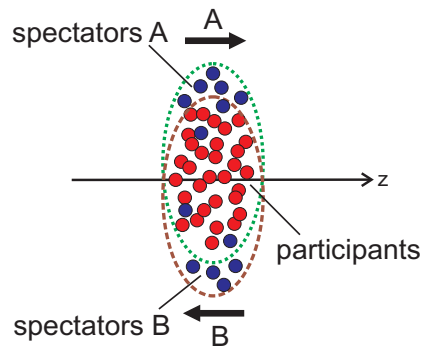}
\caption{Decomposition of system into spectators and participants.
The
possibility for nucleons to become spectators even if they are in the overlap
region of colliding nuclei is specially illustrated. }
\label{fig:specpart}
\end{center}
\end{figure}

We consider the collision of two identical heavy nuclei with mass number $A$,
and we analyze the probability for a nucleon to become a spectator or a
participant.
The many-particle system created in heavy-ion collisions can be divided into three subsystems
(Fig.~\ref{fig:specpart}): spectators
from the projectile, (A), spectators from the target (B) and the participant
particles (P).
The conservation of
four-momentum provides us
with the following expressions for the
total energy and longitudinal momentum in the collider c.m.s.:
\begin{eqnarray}
E_{\rm tot} & = & E_A + E_B + E_P,\label{eq:Etot} \\
P^z_{\rm tot} & = & P^z_A + P^z_B + P^z_P\ =\ 0 \,.
\label{eq:Pztot}
\end{eqnarray}
The c.m. rapidity of the participant system can then be expressed as
\begin{equation}
y_P\, =\, \frac{1}{2}\ln{\frac{E_P + P^z_P}{E_P - P^z_P}}\,.
\label{eq:yparticipants}
\end{equation}
Using Eqs.~\eqref{eq:Etot}-\eqref{eq:Pztot} we can express $y_P$ in terms
of spectator energy and momentum \cite{Csernai3}
\begin{equation}
y_{c.m.} \approx y_P\ =\
\operatorname{arctanh}\left[\frac{-\left( P^z_A + P^z_B\right)}
{E_{\rm tot} - E_A - E_B}\right] \,.
\label{eq:ypgen}
\end{equation}
Next we will neglect the initial
Fermi motion of nucleons in the colliding nuclei
compared to their collective collision energy.
In this case we can express $E_{\rm tot}$, $E_{A(B)}$ and $P^z_{A(B)}$
in terms of spectator numbers $N_A$ and $N_B$ as
\begin{eqnarray}
E_{\rm tot} & = & 2 A\,p^0_{\rm i},\\
E_A & = & N_A\,p^0_{\rm i},\\
E_B & = & N_B\,p^0_{\rm i},\\
P^z_A & = & N_A\,p^z_{\rm i},\\
P^z_B & = & -N_B\,p^z_{\rm i},
\end{eqnarray}
where $p^0_{\rm i} = \sqrt{s}/2$ and $p^z_{\rm i} = \sqrt{s/4 - m_N^2}$ are
the initial nucleon energy and momentum respectively, and hence the spectator
nucleon energy and momentum.
Here $m_N = 938$~MeV/$c^2$ is the nucleon mass.
The c.m. rapidity can be expressed now in terms of the spectator numbers $N_A$
and $N_B$ as
\begin{equation}
y_P(N_A,\, N_B)\ =\ \operatorname{arctanh}\left(\frac{N_B - N_A}
{2A - N_A - N_B}\,v_{\rm i}\right) \,,
\label{eq:yNANB}
\end{equation}
where $v^{\rm i} = \displaystyle\frac{{p^z_{\rm i}}}{{p^0_{\rm i}}}$ is the
initial velocity of nucleons.
It is seen from this relation that within our model only discrete set of values
of $y^{c.m.}$ are possible.
This is a consequence of
neglecting
the Fermi motion of nucleons,
which would smear the momenta of spectators if accounted for,
and also a consequence of neglecting the interaction
between spectators and participants.
Thus, we can determine the probabilities of different participant rapidities,
$y^{c.m.}$, if we can determine the probabilities of spectator numbers
$N_A$ and $N_B$.

\subsection{Spectator number probability}
\begin{figure}
\begin{minipage}{.48\textwidth}
\centering
\includegraphics[width=\textwidth]{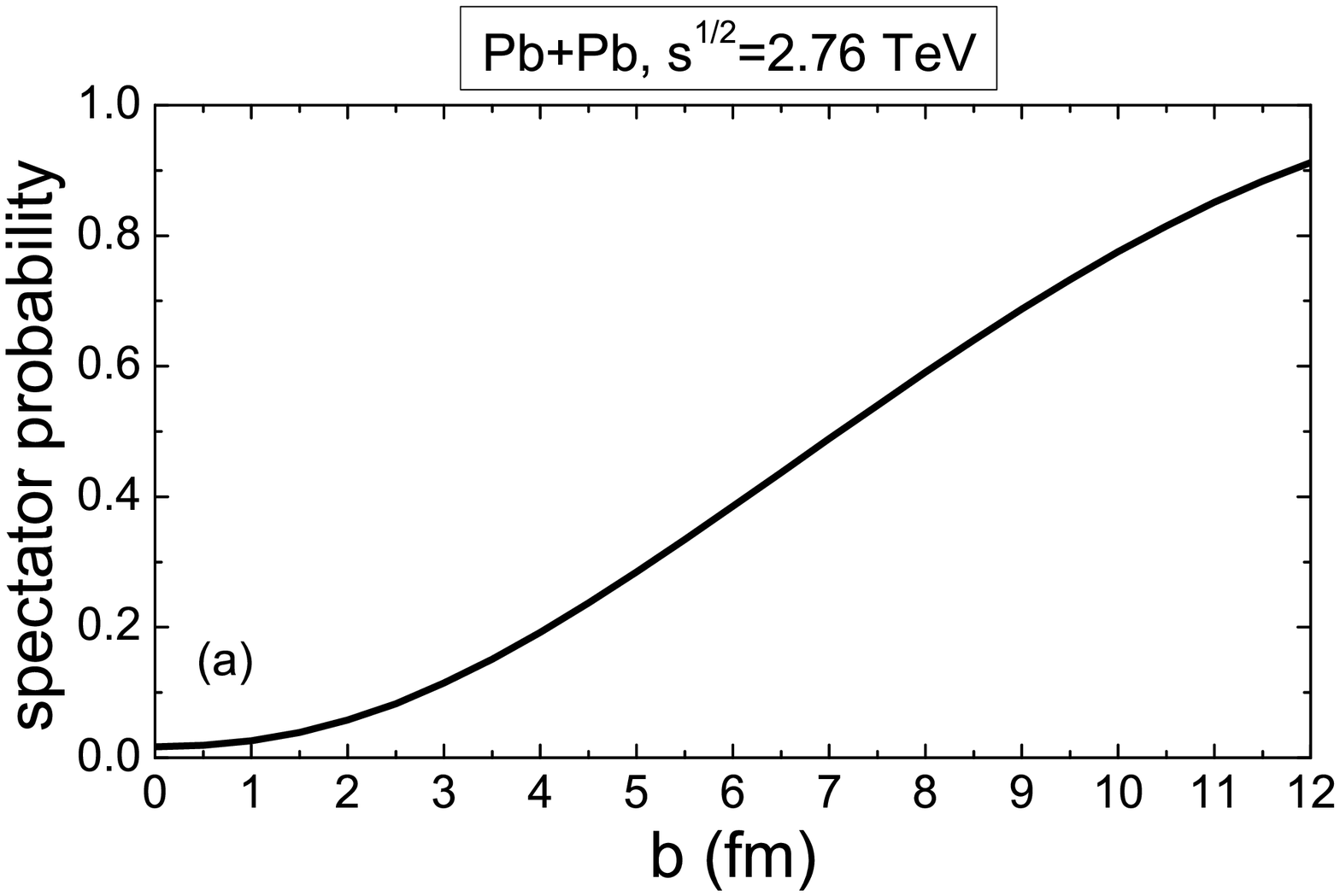}
\end{minipage}
\begin{minipage}{.48\textwidth}
\includegraphics[width=\textwidth]{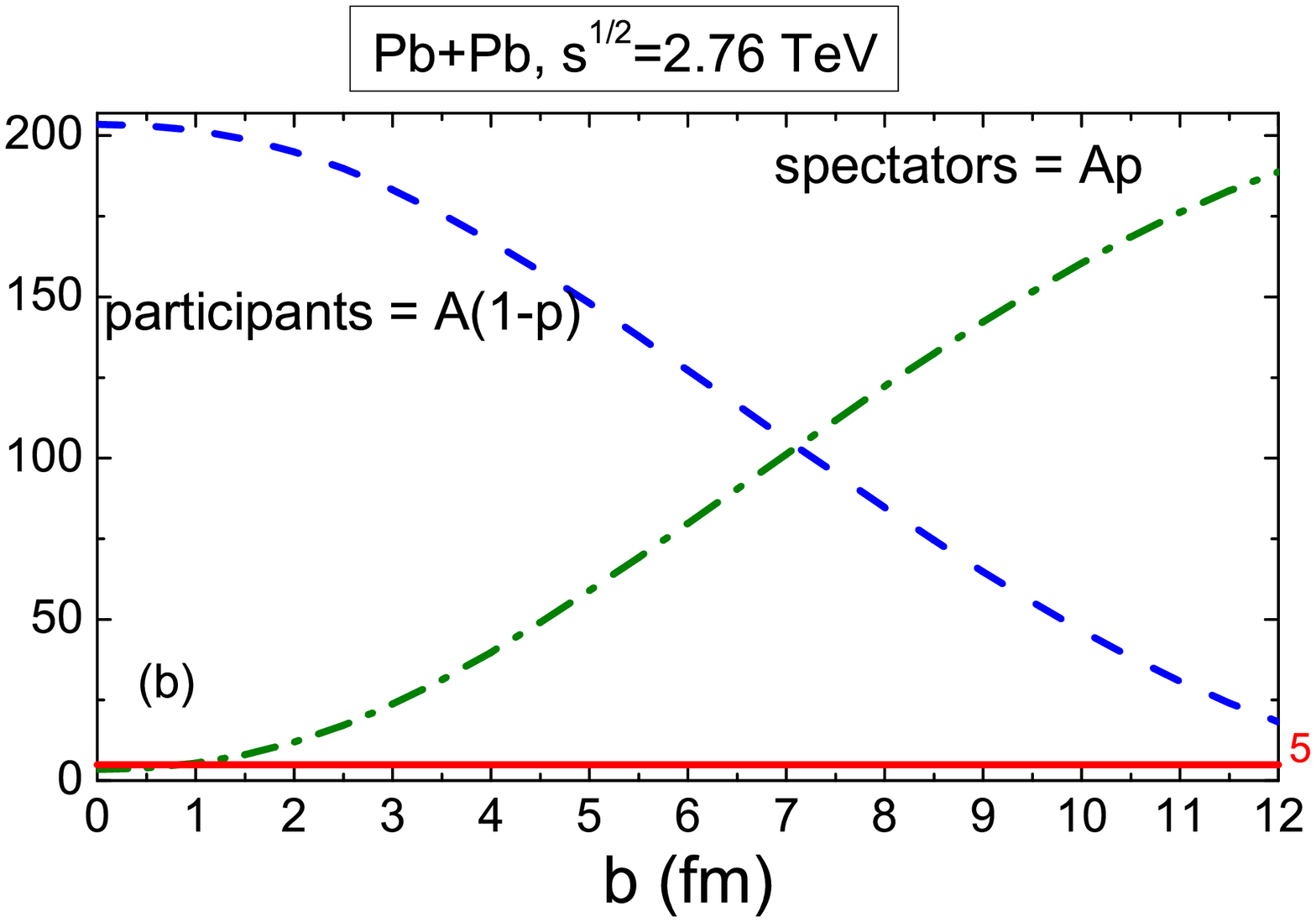}
\end{minipage}
\caption{(a)
The nucleon spectator probability dependence
on impact parameter for Pb+Pb collisions at $\sqrt{s_{NN}}=2.76$~TeV.
(b)
The same dependence of the average number of spectators
(dotted green line) and participants (dashed blue line).
The solid red line indicates the lower threshold value
for these numbers, which determines the conditions
when the Gaussian approximation of Poisson
distribution is applicable.}
\label{fig:PbPb2_76TeVprob}
\end{figure}

The transverse distribution of spectators in the collision of
heavy ions can be evaluated from the Glauber-Sitenko approach
\cite{Glauber,Sitenko,GlauberModelReview}
\begin{eqnarray}
& & \frac{d^2N_{\rm spec}}{dxdy} =  T_A(x-b/2,y) \,
\left[1 - \frac{\sigma_{NN} T_B(x+b/2,y)}{A}\right]^A \nonumber \\
&  & \qquad + T_B(x+b/2,y) \, \left[1 - \frac{\sigma_{NN} T_A(x-b/2,y)}{A}\right]^A,
\label{eq:specdis}
\end{eqnarray}
where $b$ is the impact parameter, $\sigma_{NN}$ is the nucleon-nucleon reaction
cross section and
\[T_{A(B)}(x,y)\, =\, \int dz \, \rho_{A(B)} (x,y,z)\]
is the thickness function of the projectile (target) nucleus.
Here
\begin{equation}
\rho_{A(B)} (x,y,z)\, \propto\, \left[ 1 + \exp\left(\frac{r-R}{\alpha}\right)
\right]^{-1}
\label{eq:rho}
\end{equation}
is the Woods-Saxon nuclear density distribution in nucleus.
For large mass number, $A$, we have
$\left(1 - \sigma_{NN} T_B/ A \right)^A
\approx \exp\left( - \sigma_{NN} T_B\right)$
and in this case Eq.~\eqref{eq:specdis} is often written in terms of exponents.
The first term on the right-hand side of Eq.~\eqref{eq:specdis} is the
transverse distribution of the
spectators from the projectile nucleus and the second term
is the
transverse distribution of the spectators from the target nucleus.
The probability that a nucleon from the projectile will become
a spectator
(which is the same as the probability for
a nucleon from the target due to symmetry)
can be expressed as
\begin{eqnarray}
p_A = p_B & = &
p =
\frac{1}{A}\, \int dx\,dy\, T_A\left(x - b/2, y\right) \nonumber \\
& & \times \left(1 - \frac{\sigma_{NN} T_B(x+b/2,y)}{A}\right)^A.
\label{eq:psingle}
\end{eqnarray}
The dependence of this probability on the impact parameter
for Pb+Pb collisions at $\sqrt{s_{NN}}=2.76$~TeV is
depicted in Fig.~\ref{fig:PbPb2_76TeVprob}.
The values of parameters used for calculations
are $\sigma_{NN} = 70$~mb \cite{pdg,Antchev:2011vs},
$A=208$, $R=6.53$~fm and $\alpha=0.545$.
Using parameter $p$ from \eqref{eq:psingle} we can determine the
probability that there will be $N_{A(B)}$
spectators in the projectile (target) as a binomial distribution
\begin{eqnarray}
p(N_A) & = & \binom{A}{N_A}\, p^{N_A}\, (1-p)^{A-N_A}\,, \\
p(N_B) & = & \binom{A}{N_B}\, p^{N_B}\, (1-p)^{A-N_B} \,.
\label{eq:binomial}
\end{eqnarray}
Here we have assumed that the initial many-nucleon distribution
function can be approximately expressed as a product
of one-nucleon distribution
functions, i.e. the momenta and spatial positions of nucleons are uncorrelated.
Next, one can take the number of spectators
in the projectile as independent
of the number of spectators in the target.
This is not exactly true: e.g. if there are
participants from one nucleus
then there were reactions between the colliding
nucleons, and there should also be
participants from the other nucleus.
This implies that, for a fixed number, $N_A$, of spectators in projectile nucleus
we can expect the number of spectators, $N_B$, in target nucleus
to fluctuate around the value $\langle N_B \rangle \approx N_A$.
Analogous statement can be found in Ref. \cite{Gorenstein2008}
where this subject was analyzed within the microscopic transport models.
So, the numbers of spectators from different nuclei are not
fully uncorrelated.
Meanwhile, for the sake of simplicity,
we assume that the number of spectators
in the projectile is independent
of the number of spectators in the target.
But we can expect this approximation to
work well if we have colliding heavy
ions with large mass
numbers.
Using this approximation we can write
\begin{equation}
p(N_A,\,N_B)\, \approx\, p(N_A)\, p(N_B) \,.
\label{eq:pNANB}
\end{equation}
Using this probability one can then determine the distribution function of
the corresponding c.m. rapidity of the participants, see Eq.~\eqref{eq:yNANB}.

\subsection{Gaussian approximation and rapidity distribution}

As mentioned above, in our approach $y^{P}$ takes discrete set
of values. Because we neglect the smearing of momentum
of spectators, the rapidity is defined solely by the spectator
numbers, $N_A$ and $N_B$, which so far take discrete sets of values.
It is possible to obtain a continuous rapidity
distribution if we allow the quantities $N_A$ and $N_B$
to take continuous values. It is well known that for
some conditions the binomial
distributions $p(N_A)$ and $p(N_B)$ can be accurately approximated
by the Gaussian distribution 
with mean $Ap$ and variance $Ap(1-p)$ as
\begin{equation}
p(N_{A(B)}) \Rightarrow
\rho(N_{A(B)}) =
\frac{\exp\left(-\frac{(N_{A(B)}-Ap)^2}{2 Ap(1-p)}\right)}{\sqrt{2\pi Ap(1-p)}}.
\label{eq:gauss}
\end{equation}
In our case these conditions are the following: the average spectator
and participant numbers $Ap$ and $A(1-p)$ are large enough, e.g.
$Ap>5$ and $A(1-p)>5$.
It is seen from Fig.~\ref{fig:PbPb2_76TeVprob} that
these conditions are quite well satisfied in our model
for heavy ions, especially for non-central collisions.
Using the Gaussian approximation \eqref{eq:gauss} we can write the
rapidity distribution function as
\begin{eqnarray}
f_P(y)\, & = & \, \int_{0}^{A} dN_A\,
\int_{0}^{A} dN_B\, \rho(N_A)\, \rho(N_B) \nonumber \\
& & \times \, \delta\big[\, y - y_P(N_A,N_B)\, \big] \nonumber \\
& = & \, \int_{-\infty}^{\infty} dN_A\,
\int_{-\infty}^{\infty} dN_B\, \tilde{\rho}(N_A)\,
\tilde{\rho}(N_B) \nonumber \\
& & \times \, \delta\big[\, y - y_P(N_A,N_B)\, \big],
\label{eq:df1}
\end{eqnarray}
where to switch to infinite
integration limits we introduce
\begin{equation}
\tilde{\rho}(N_{A(B)}) \, = \,
\rho(N_{A(B)}) \, \operatorname\theta(A-N_{A(B)}) \,
\operatorname\theta(N_{A(B)}).
\label{eq:tilde}
\end{equation}
Let us now define the new variables:
\begin{equation*}
N\, =\, \frac12 \, (N_A + N_B) \, , \quad n\, =\, N_B - N_A \,,
\end{equation*}
\begin{eqnarray}
f_P(y) &=& \int_{-\infty}^{\infty} dN\, \int_{-\infty}^{\infty} dn\,
\tilde{\rho}(N+n/2)\, \tilde{\rho}(N-n/2) \nonumber \\
& & \quad \times \delta\left[\, y - \operatorname{arctanh}\left(\frac{n}
{2(A - N)}\,v_{\rm i}\right)\, \right] \,.
\label{eq:df3}
\end{eqnarray}
Then we make a transformation to a new variable in the
$\delta$-function in accordance with the rule:
$\delta[y-f(n;N)] = \delta[n-F(y;N)]/|f'(\overline{n})| $,
where
\begin{equation}
\overline{n}\, =\, F(y;N)\, =\,  \frac 1{v_{\rm i}}\, 2(A - N)\, \tanh{y}\,.
\end{equation}
After introducing
$f'(\overline{n})$
explicitly, the rapidity distribution becomes
\begin{equation}
f_P(y)\, =\,
\int_{-\infty}^{\infty} dN\,
\tilde{\rho}(N+\overline{n}/2)\, \tilde{\rho}(N-\overline{n}/2)\,
\frac{2(A - N)}{v_{\rm i} \cosh^2{y}} \,.
\label{eq:df15}
\end{equation}
In order to compute the integral in \eqref{eq:df15} we will use
the following approximation
\begin{equation}
\tilde{\rho}(N\pm\overline{n}/2) \approx \rho(N\pm\overline{n}/2).
\end{equation}
This approximation works well in the case when
the original binomial
distribution \eqref{eq:binomial} is well
approximated by the Gaussian \eqref{eq:gauss}.
The presence of the Gaussian allows one to neglect the Heaviside theta functions
in the integration in \eqref{eq:df15},
which we perform using
expression \eqref{eq:gauss} for $\rho(N\pm\overline{n}/2)$ and obtain
\begin{equation}
f_P(y)\, =\, \sqrt{\frac{A(1-p)}{\pi p}} \,
\frac{v_{\rm i}^2 \,
\exp\left[-\frac{A(1-p)}{p}\, \frac{\tanh^2{y}}{v_{\rm i}^2 + \tanh^2{y}}
\right]}{\cosh^2{y}
\left[v_{\rm i}^2 + \tanh^2{y}\right]^{\frac{3}{2}}}.
\label{eq:fpy-final}
\end{equation}

\begin{figure}
\begin{center}
\includegraphics[width=0.48\textwidth]{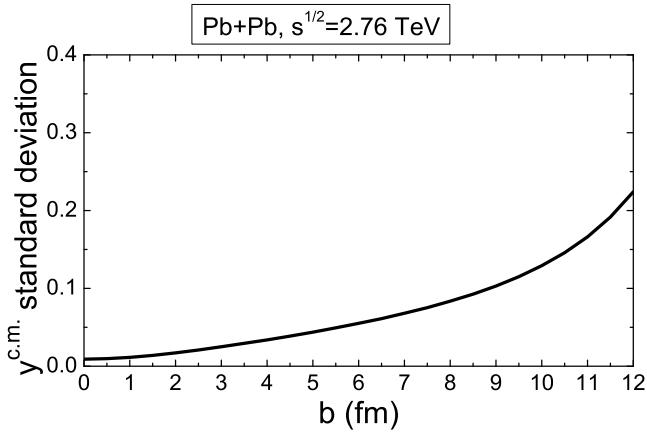}
\caption{The dependence of the standard deviation, $\sqrt{\delta y^2}$,
on the impact parameter. The calculations
are made for Pb+Pb collisions at
$\sqrt{s_{NN}}=2.76$~TeV.}
\label{fig:PbPb2_76TeVstdev}
\end{center}
\end{figure}

\subsection{Ultra-relativistic limit and distribution at mid-rapidity}
It is useful to analyze Eq.~\eqref{eq:fpy-final} in the ultra-relativistic
limit, i.e. when $v_{\rm i} \to 1$. We obtain
\begin{equation}
f^{\rm UR}_P(y)\, =\, \sqrt{\frac{A(1-p)}{\pi p}} \, \frac{
\exp\left[-\frac{A(1-p)}{p}\,
\frac{\tanh^2{y}}{1 + \tanh^2{y}}\right]}
{\cosh^2{y} \left[1+\tanh^2{y}\right]^{\frac{3}{2}}} \,.
\label{eq:fpy-final-UR}
\end{equation}

If we now consider collisions of identical nuclei at
fixed impact parameter for different
collision energies then we can see that the
only parameter in Eq.~\eqref{eq:fpy-final-UR}
which depends on collision energy is the single-nucleon
spectator probability $p$. It depends on collision
energy only due to possible
energy dependence of the nucleon-nucleon cross
section $\sigma_{NN}$, see Eq.~\eqref{eq:psingle}.
It is known that for a wide energy range
(e.g. for SPS and RHIC energies) $\sigma_{NN}$
depends very weakly on collision energy and
therefore we can claim that under such conditions
the participant center-of-mass rapidity
distribution is invariant of collision energy.

Another limit, which can be explored is the
distribution at mid-rapidity, i.e.
around $y=0$.
We can expect that the c.m. rapidity fluctuations in heavy-ion collisions
should be quite small
(and subsequent calculations in next section seem to confirm this)
and for small
c.m. rapidity values $y^{c.m.}$ we can approximate
hyperbolic functions in Eq.~\eqref{eq:fpy-final} as
$\cosh{y} \approx 1$ and $\tanh{y} \approx y$.
Also, since we deal with relativistic collision energies
$v_{\rm i}$ should be close to 1, which allows us to write
$(\tanh{y})^2 + v_{\rm i}^2 \approx v_{\rm i}^2$.
With these approximations the rapidity distribution becomes
\begin{equation}
f_P(y)\, =\, \sqrt{\frac{A(1-p)}{\pi p \, v_{\rm i}^2}} \,
\exp\left[-\frac{A(1-p)}{p\,v_{\rm i}^2}\, y^2\right] \,.
\label{eq:fpy-gauss}
\end{equation}
This is actually a Gaussian distribution around $y=0$ with variance
\begin{equation}
\delta y^2\ =\ \displaystyle \frac{p\, v_{\rm i}^2}{2A(1-p)} \,.
\label{eq:fpy-variance}
\end{equation}
The expression for the variance gives
the
following result: rapidity fluctuations
are stronger for higher nucleon spectator probability $p$,
i.e. they are increasing
with the increase of impact parameter.
For $p = 0$ there are no spectators in the system and therefore
collider c.m.s. and participant c.m.s. coincide.
This result is reproduced by Eq.~\eqref{eq:fpy-gauss} in our model.

The dependence of standard deviation, $\sqrt{ \delta y^2 }$, given by this
Gaussian distribution on collision impact parameter for Pb+Pb collisions at
$\sqrt{s}=2.76$~TeV is depicted in Fig.~\ref{fig:PbPb2_76TeVstdev}.
It is seen that the standard deviation of rapidity (which is basically a
distribution width) stays significantly smaller than 1, hence justifying
our approximation of rapidity distribution for small values of $y$.
We can expect the total rapidity distribution given by the Gaussian in
Eq.~\eqref{eq:fpy-gauss} to work well for most conditions in heavy-ion
collisions.

\begin{figure}[!t]
\begin{center}
\includegraphics[width=0.48\textwidth]{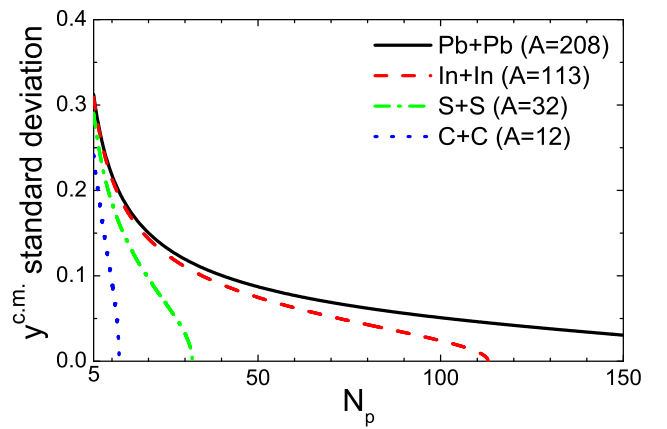}
\caption{The dependence of standard deviation,
$\sqrt{\delta y^2}$, see Eq. (\ref{eq:fpy-variance-Np}),
of the rapidity distribution given by the Gaussian
\eqref{eq:fpy-gauss} on the average number
of participants, $N_p$, for different colliding nuclei,
Pb+Pb, In+In, S+S, C+C.}
\label{fig:stdev-Np}
\end{center}
\end{figure}

Expression \eqref{eq:fpy-variance} for $\sqrt{ \delta y^2 }$ can
be rewritten in terms of mass number $A$ and average number of
participants, $N_p = A(1-p)$. It reads as
\begin{equation}
\delta y^2\ =\  \displaystyle \frac{v_{\rm i}^2}{2}\,\left(\frac{1}{N_p}-\frac{1}{A}\right).
\label{eq:fpy-variance-Np}
\end{equation}
It is interesting to explore the dependence of rapidity fluctuations on the
average number of participants $N_p$ for different pairs of colliding
nuclei $A+A$.
This dependence is depicted in Fig.~\ref{fig:stdev-Np}
for Pb+Pb, In+In, S+S and C+C collisions.
There we take the initial nucleon velocity, $v_{\rm i}=1$, since just
ultra-relativistic collision energies are considered.
Notice that,
for fixed average number of participants, $N_p$,
the rapidity fluctuations are stronger
in collisions of heavier nuclei. For instance, if we consider central
collisions of light nuclei, then the rapidity fluctuations in
``equivalent'' non-central collisions of heavier nuclei will be bigger.
Here ``equivalent'' means that in both colliding systems the average number
of participants, $N_p$, is the same. Similar
amplification of fluctuations with respect to the mass number
was obtained in Ref.~\cite{Gorenstein2008}.

\section{Calculation results}

Let us calculate the participant c.m. rapidity distribution
for various collision conditions. The dependence of nucleon spectator probability $p$ from \eqref{eq:psingle}
on the impact parameter was considered in section \ref{sec:model}.
This dependence for Pb+Pb collisions at $\sqrt{s}=2.76$~TeV is depicted in Fig.~\ref{fig:PbPb2_76TeVprob}a.
It is seen
that the
nucleon spectator probability
strongly depends on the centrality of the collision: it is small,
but non-zero for central collisions, about 0.5 for mid-central collisions,
and closer to unity for peripheral collisions.

\begin{figure}
\begin{center}
\includegraphics[width=0.48\textwidth]{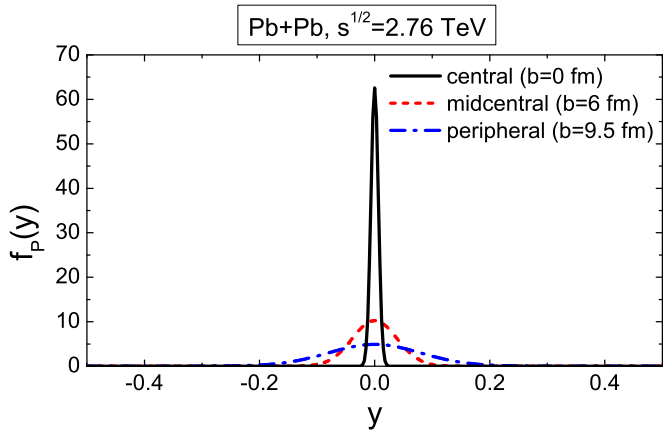}
\caption{Participant center-of-mass rapidity distribution
for Pb+Pb collisions
at $\sqrt{s_{NN}}=2.76$~TeV
for three different centralities.}
\label{fig:PbPb2_76TeVthree}
\end{center}
\end{figure}

Next, we will explore the participant c.m.
rapidity distribution for different centralities but
for the same collision energy using Eq.~\eqref{eq:fpy-final} for calculations.
The rapidity distributions for Pb+Pb collisions at $\sqrt{s}=2.76$~TeV
are depicted in
Fig.~\ref{fig:PbPb2_76TeVthree} for three different centralities:
central ($b=0$~fm),
mid-central ($b=6$~fm) and peripheral ($b=9.5$~fm).
It is seen that the  rapidity distribution, $f_P(y)$,
depends strongly on the impact parameter, just as the
nucleon spectator probability.
At small impact parameter the $y^{c.m.}$ fluctuations are
small and appear to be insignificant.
However, increasing the impact parameter up
to 9.5~fm (peripheral collisions) results
in a significant increase of the $f_P(y)$-distribution width
compared to central collisions.
So, first of all in peripheral collisions
the c.m. rapidity fluctuations may play an important
role when calculating different measurable
rapidity distributions
\cite{Csernai3}.
As was discussed, the rapidity fluctuations
are smaller in central collisions of light nuclei
compared to the fluctuations in non-central collisions
of heavier nuclei when the number of participants is the same
in both cases.
It was also checked that in all three cases depicted
in Fig.~\ref{fig:PbPb2_76TeVthree} the
calculated rapidity distribution virtually coincides
with the Gaussian distribution given by \eqref{eq:fpy-gauss}.

\begin{figure}
\begin{center}
\includegraphics[width=0.48\textwidth]{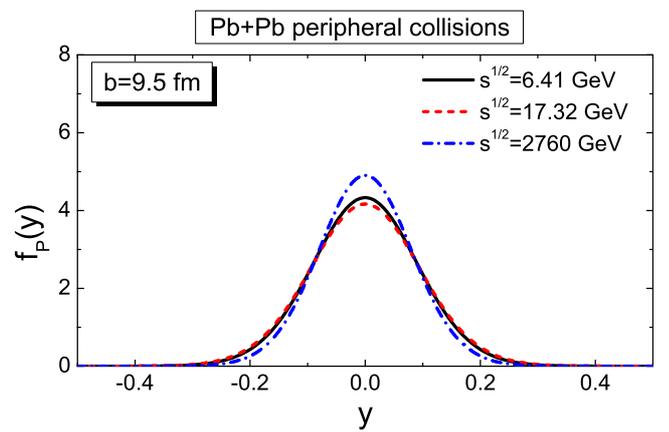}
\caption{Participant center-of-mass
rapidity distribution for peripheral Pb+Pb collisions at
$\sqrt{s_{NN}}=6.41$~GeV ($E_{\rm kin}=20$~GeV),
$\sqrt{s_{NN}}=17.32$~GeV ($E_{\rm kin}=158$~GeV) and
$\sqrt{s_{NN}}=2.76$~TeV.}
\label{fig:PbPbenerscan}
\end{center}
\end{figure}

It is interesting to explore the influence of the collision energy on
the participant c.m. rapidity fluctuations. To do that we consider
Pb+Pb collisions at three different energies:
$\sqrt{s_{NN}}=6.41$~GeV ($E_{\rm kin}=20$~GeV),
$\sqrt{s_{NN}}=17.32$~GeV ($E_{\rm kin}=158$~GeV) and
$\sqrt{s_{NN}}=2.76$~TeV. The first
two energies correspond to CERN-SPS experiments
and the third one to the CERN-LHC experiment.
We take $\sigma_{NN} = 33$~mb for both SPS
energies~\cite{pdg} and $\sigma_{NN} = 70$~mb for LHC energy.
The calculation results for different energies
for peripheral collisions ($b=9.5$~fm) are presented
in Fig.~\ref{fig:PbPbenerscan}.

We can see that the collision energy
influence on participant c.m. rapidity fluctuations
is rather weak, especially compared to the centrality dependence.
This can be explained by the fact that for high energies the rapidity
distribution is well described in the ultra-relativistic
limit \eqref{eq:fpy-final-UR},
and the difference between LHC energy and SPS energies is due to
doubling of the
nucleon-nucleon cross section which still does not lead to a
significant change in rapidity fluctuations.

\subsection{Other Theoretical Estimates}

Longitudinal fluctuations arising from initial
state fluctuations in the PACIAE parton and hadron
molecular dynamics model were analyzed recently
\cite{Cheng11},
and the fluctuation of the center-of-mass rapidity of the system was
conservatively estimated to be $\Delta y^{c.m.} = 0.1$, by
neglecting all pre-equilibrium emission effects that increase the
$y^{c.m.}$-fluctuations.

The unique separation of participants and spectators in realistic
situations is not trivial. Between the participants and spectators
some level of interaction may remain at the separation, and a small number
of nucleons cannot be classified definitely as either participants or
spectators. This indefiniteness may
result in increase or decrease
of
$y^{c.m.}$-fluctuations. Longitudinal fluctuations may influence other
observables also
\cite{HP11,LP12}.

\begin{figure}
\begin{center}
\includegraphics[width=0.48\textwidth]{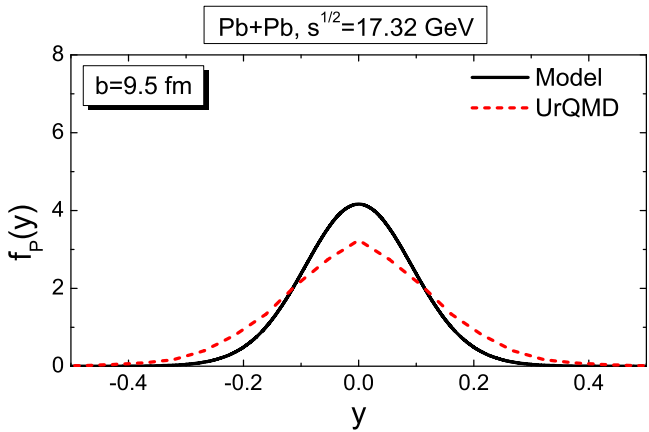}
\caption{Participant center-of-mass rapidity
distribution for Pb+Pb peripheral collisions
at $\sqrt{s_{NN}}=17.32$~GeV ($E_{\rm kin}=158$~GeV)
within the present model and in the UrQMD model.}
\label{fig:PbPb158UrQMD}
\end{center}
\end{figure}

Next we will compare our model calculations
with corresponding calculations within
the UrQMD microscopic transport model~\cite{UrQMD1998,UrQMD1999}.
The EbE c.m. rapidity
can be computed in UrQMD using Eq.~\eqref{eq:yparticipants}.
There we can account for the Fermi motion as well as for initial
nucleon correlations, however, a large number of simulated UrQMD
events are necessary to obtain smooth distribution.
The comparison of the participant c.m.
rapidity distribution calculated within our model
and with the UrQMD model for Pb+Pb collisions at
$\sqrt{s_{NN}}=17.32$~GeV ($E_{\rm kin}=158$~GeV)
with $\sigma_{NN}=33$~mb
for peripheral collisions ($b=9.5$~fm)
is presented in Fig.~\ref{fig:PbPb158UrQMD}.

One can see that there is a difference in the distributions
calculated within these two models, the distribution
from UrQMD is wider.  The difference
is relatively small compared to the difference
arising from changing the collision centrality
(see Fig.~\ref{fig:PbPb2_76TeVthree}).
The difference, which is seen in Fig.~\ref{fig:PbPb158UrQMD} can be
attributed
to neglecting
the initial many-nucleon
correlations as well as the spectator number
correlations for nucleons from
the colliding nuclei, which were assumed
in our model.
It could also be questioned whether nucleons, which did not take part
in any reaction in UrQMD may be correctly identified as spectators in
the Glauber-Sitenko approach.

\begin{figure}
\begin{minipage}{.48\textwidth}
\centering
\includegraphics[width=\textwidth]{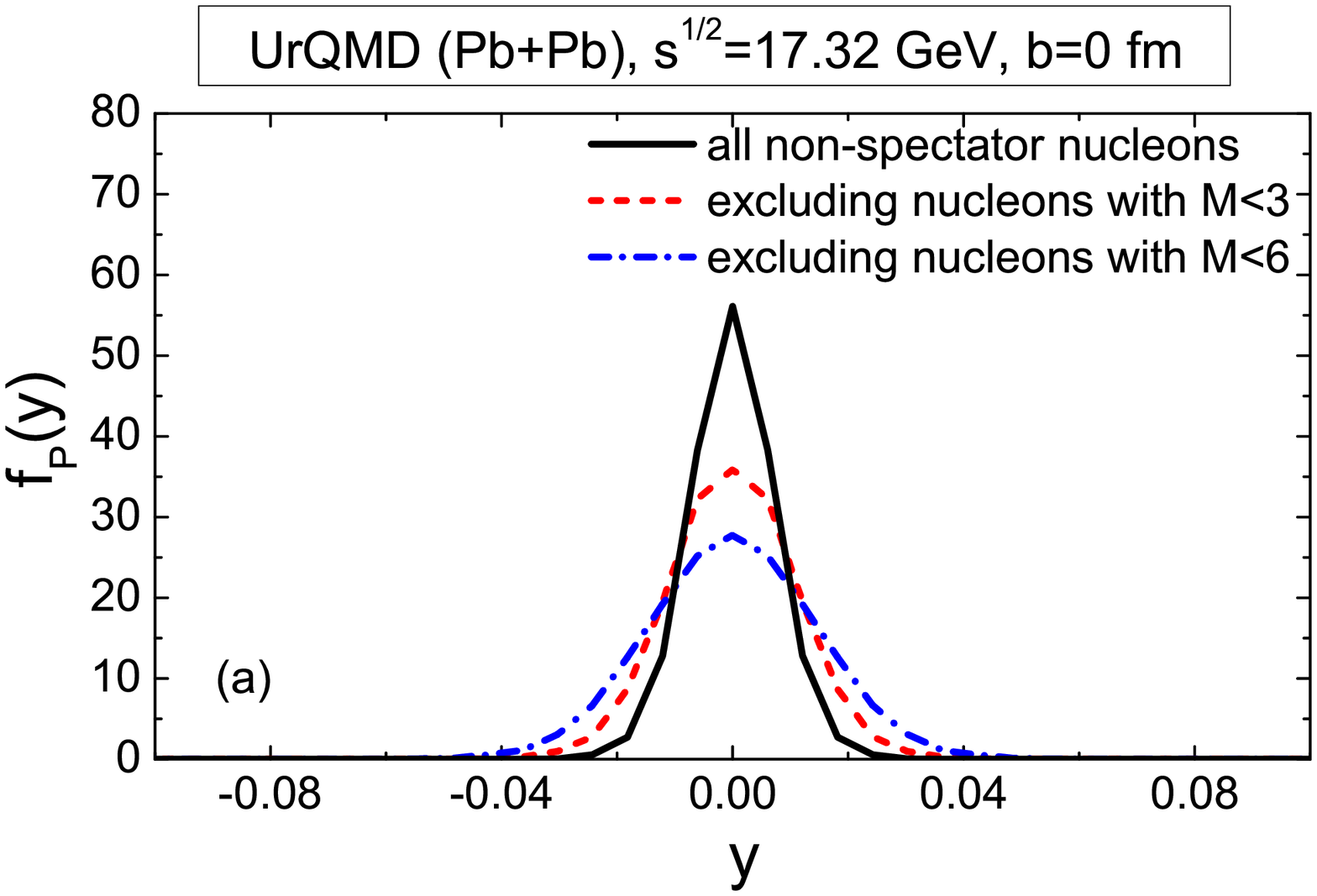}
\end{minipage}
\begin{minipage}{.48\textwidth}
\includegraphics[width=\textwidth]{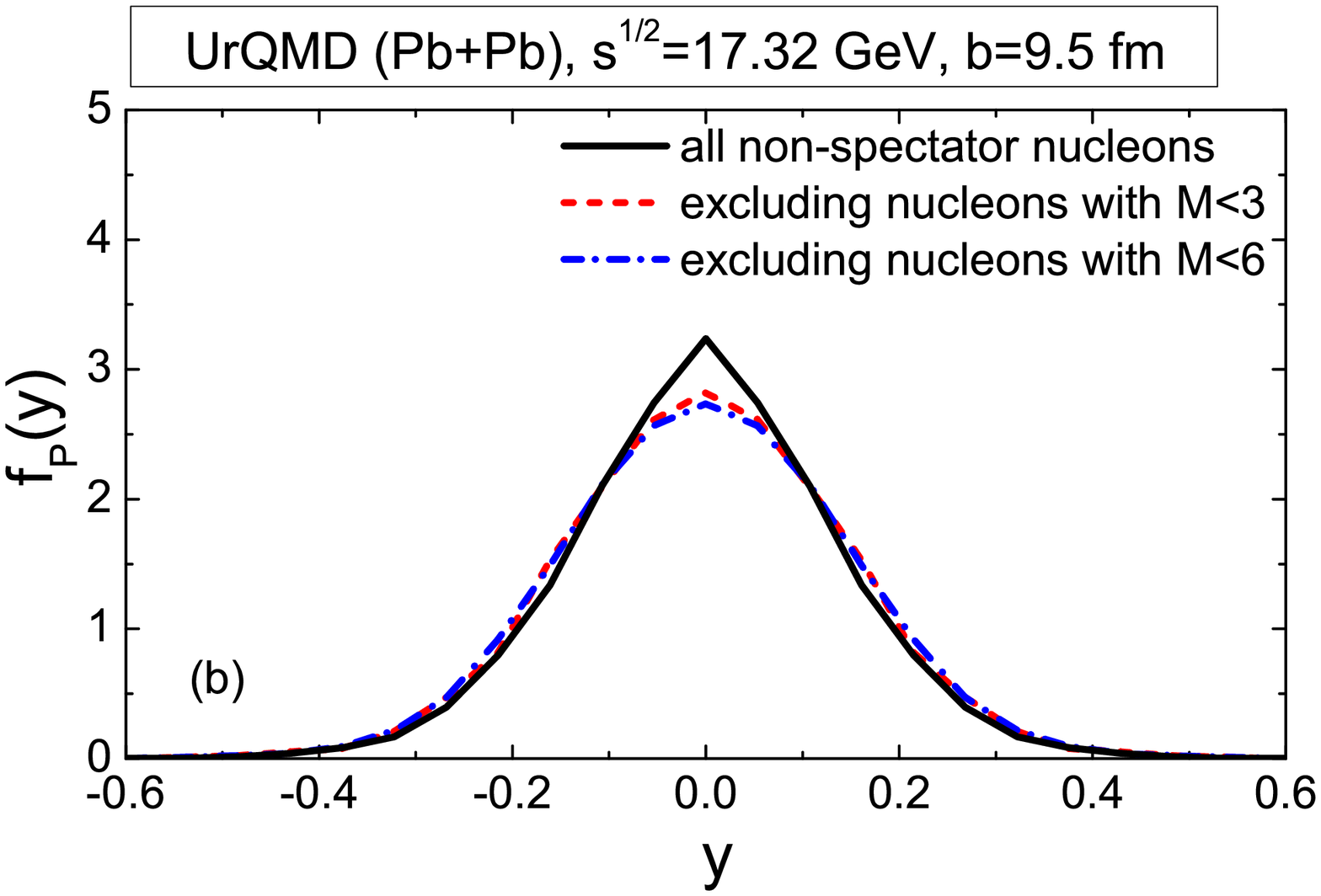}
\end{minipage}
\caption{Participant center-of-mass
rapidity distribution for (a) central and (b) peripheral Pb+Pb
collisions calculated in the
UrQMD model for different participant definitions, where nucleons
which collided less than the given limit are excluded from the participants.}
\label{fig:diff-M}
\end{figure}

The separation of spectators from participants is studied in Ref.
\cite{ANC13}.
Here the pre-equilibrium emission of one or two nucleons plays a
non-negligible role. The (thermal) equilibration is demonstratively
not present for particles,
which interacted fewer than 4-6 times. These
cannot be considered as parts of a participant system, and usually have
large longitudinal and small transverse momenta, although these do
not reach the Zero Degree Calorimeters, so experimentally these are
not identified as spectators.
Similar considerations were used to describe the strangeness enhancement
within the core-corona picture~\cite{CoreCorona}, where nucleons
which have scattered only once were regarded as corona nucleons and were not
part of a fireball.
In central and semi-peripheral
reactions these pre-equilibrium particles may influence the
$y^{c.m.}$-fluctuations considerably. For example, if we exclude
nucleons from the participant system which collided fewer than six times,
$M < 6$, then in central collisions
the center-of-mass rapidity fluctuation doubles,
see Fig.~\ref{fig:diff-M}a. However, there is little change
in rapidity distribution in case of peripheral collisions
(Fig.~\ref{fig:diff-M}b).

\subsection{Participant angular momentum}
The model also provides an estimate of the total angular momentum
of the initial participant system.
The angular momentum, $L_{\rm tot}^P$, of the participant system can be calculated
as the difference of total angular momentum, $L_{\rm tot}$, and
the angular momentum of spectators, $L_{\rm tot}^S$. The quantities
$L_{\rm tot}$ and $L_{\rm tot}^S$ can be calculated with the use
of nuclei thickness functions $T_{A(B)}(x,y)$ and using
the transverse distribution of spectators $T_{A(B)}^S (x,y)$ as
\begin{eqnarray*}
L_{\rm tot} & = & p^z_{\rm i} \int dx dy \, x \left[ T_A(x-b/2,y) - T_B(x+b/2,y) \right], \\
L_{\rm tot}^S & = & p^z_{\rm i} \int dx dy \, x \left[ T_A^S(x,y) - T_B^S(x,y) \right], \\
L_{\rm tot}^P & = & L_{\rm tot} - L_{\rm tot}^S.
\end{eqnarray*}
The transverse distribution of spectators can be determined from the Glauber-Sitenko
model as \eqref{eq:specdis}. Another approach is to consider as participants all nucleons in the overlap
region of colliding nuclei \cite{Csernai1,Csernai2,Becattini2008,Gao2008}.

\begin{figure}
\begin{minipage}{.48\textwidth}
\centering
\includegraphics[width=\textwidth]{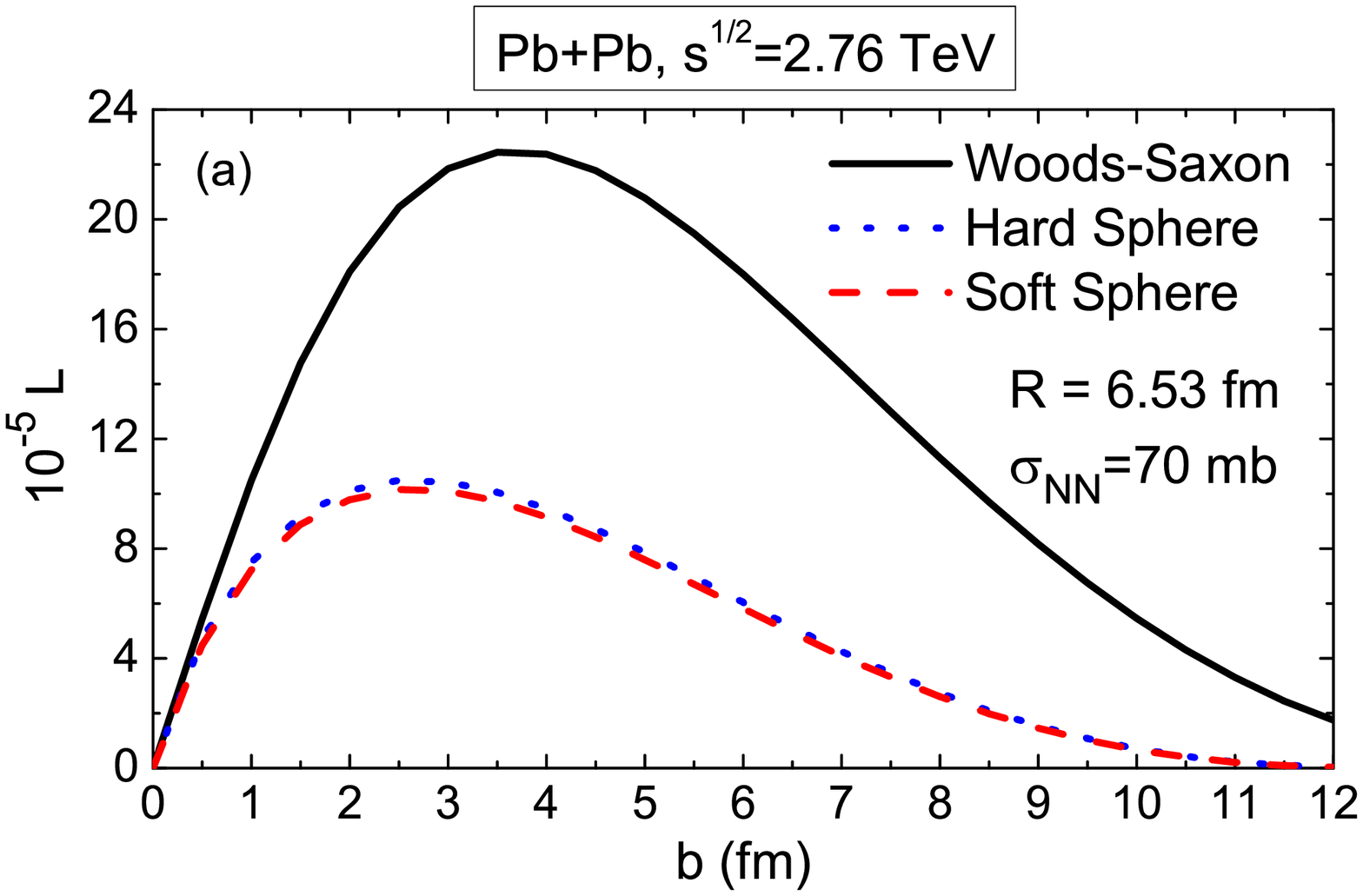}
\end{minipage}
\begin{minipage}{.48\textwidth}
\includegraphics[width=\textwidth]{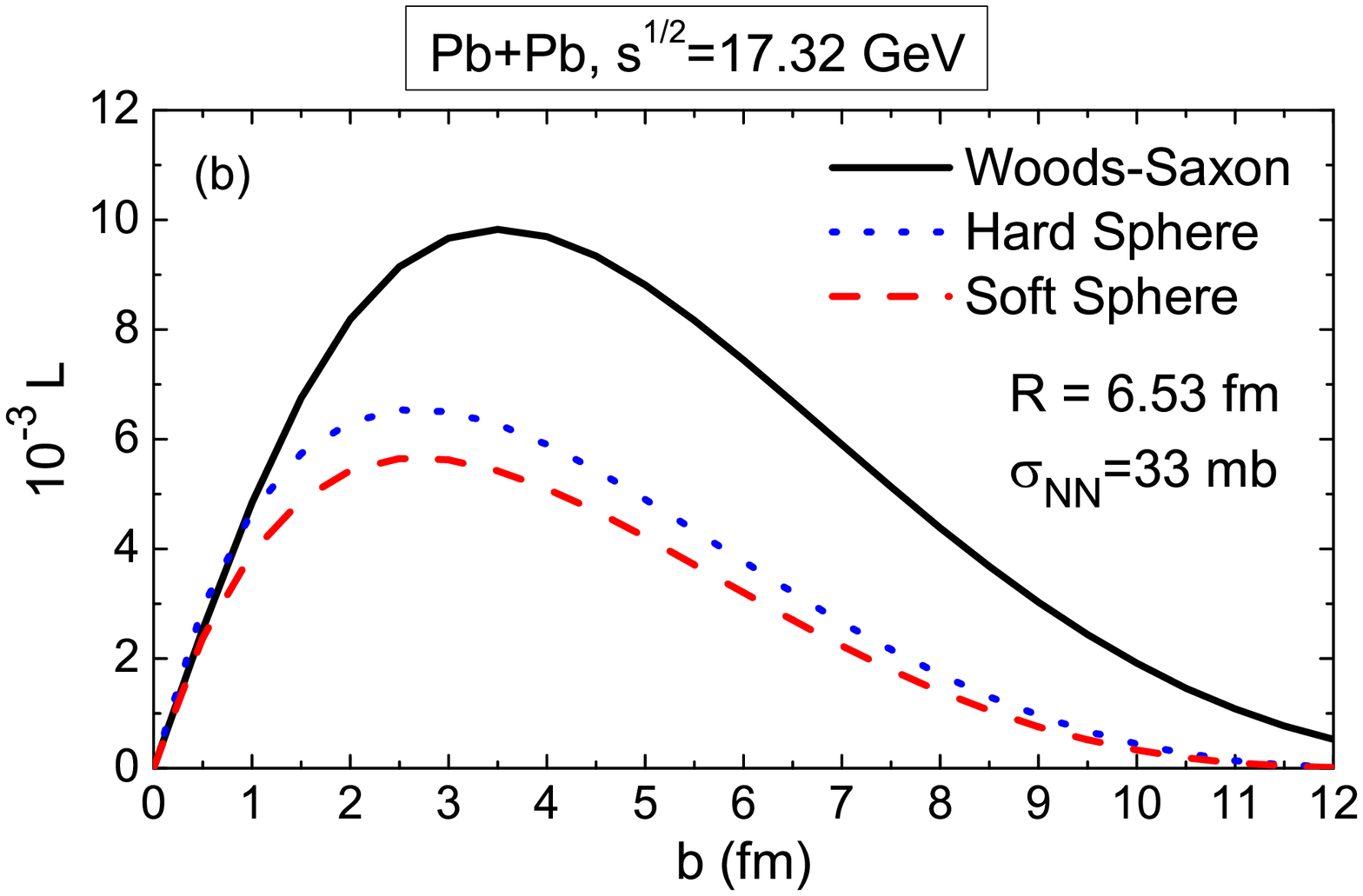}
\end{minipage}
\caption{The dependence of total angular momentum
of the participant system on impact parameter
in Pb+Pb collisions
for (a) LHC and (b) SPS conditions
for different
nuclear density profiles.}
\label{fig:angmom}
\end{figure}

First of all the angular momentum
for LHC Pb+Pb reactions at $\sqrt{s_{NN}}=2.76$~TeV
is about two orders of magnitude larger than
that at SPS energy of at $\sqrt{s_{NN}}=17.32$~GeV (see Fig.~\ref{fig:angmom})
and one order of magnitude larger than
for Au+Au reactions RHIC energy of  $\sqrt{s_{NN}}=200$ GeV
\cite{Gao2008}.

The angular momentum is the largest for nuclei with a
Woods-Saxon radial density profile, see Eq.~\eqref{eq:rho}, due to the presence of the
diffusion zone with a tail, which effectively increases angular momentum.
For nuclei with homogeneous nuclear density where the density profile has a sharp boundary
we consider all nucleons from the overlap region as participants and
all other nucleons as spectators (Hard Sphere Nuclei).
In this case the angular momentum is about a factor of two less
than for a Woods-Saxon profile (see Fig.~\ref{fig:angmom}).

If, in addition, a transparency in the overlap region is assumed due to the finite
NN cross section (Soft Sphere Nuclei)
then the angular momentum is further reduced by 2\% and 15\% at
LHC and SPS energies, respectively (see Fig.~\ref{fig:angmom}).

Thus, in fluid dynamical and in molecular dynamics models,
the assumed initial state leaves some
freedom for the angular
momentum of the participant system.

\section{Conclusions}
A simple model to calculate the participant c.m. rapidity distribution
is developed
and used to analyze the rapidity fluctuations for
different conditions in heavy-ion collisions.
In the model a weak initial nucleon-nucleon
correlation in colliding nuclei and
weak correlations between spectator numbers from different nuclei
are assumed
and the interaction between spectators and participants is neglected.
The main input parameter in the model is
the probability for a nucleon to be a spectator,
which is determined from the Glauber-Sitenko approach
in the current work. Different models
for calculating this probability are applicable.

It is shown that for small rapidity values the rapidity distribution
can be well approximated by the Gaussian distribution with variance determined
by the nucleon spectator probability and by initial
nucleon velocities. The calculation results confirm
that this approximation works well in a wide
range of collision energies and centralities.

It is shown that rapidity fluctuations strongly depend on
impact parameter -- they are stronger for
more peripheral collisions
and these
fluctuations should be taken into account in
calculation and interpretation of various
rapidity-dependent observables
\cite{Csernai3}.
It is necessary to note that, if we consider collisions of two different pairs
of nuclei, for instance $A_1+A_1$ and $A_2+A_2$ with
$A_1>A_2$, where the number of participants
is the same in both collisions, then
the rapidity fluctuations are smaller in collisions $A_2+A_2$ of lighter
nuclei. Recent studies \cite{Csernai3} indicate
a possibility of experimental measurement of the
$y^{c.m.}$ fluctuations.

The collision energy dependence of rapidity fluctuations appears
to be weak. Comparison with similar
c.m. rapidity distribution calculations
within the UrQMD model shows qualitative agreement,
however, some indefiniteness in identification of spectators and participants,
for instance pre-equilibrium emission of nucleons,
may lead to extra sources of participant c.m. fluctuations, especially
at more central collisions.

\section*{Acknowledgements}
L.P. Cs. thanks the enlightening discussions and kind hospitality at
the Frankfurt Institute for Advanced Studies, where he has done part
of this work. D. A. was supported by the National Academy of
Sciences of Ukraine (project No. 0113U001092).

\end{document}